\documentclass[
        reprint,
        aps,
        prx,
        superscriptaddress, 
        a4paper,
]{revtex4-2}
\usepackage[english]{babel}
\usepackage[utf8]{inputenc}
\usepackage{amsmath, amssymb, amsthm}
\usepackage{braket}
\usepackage{bm}
\usepackage{microtype}
\usepackage{graphicx}
\usepackage{hyperref}
\usepackage[capitalise]{cleveref}
\usepackage{siunitx}
\usepackage{lipsum}
\usepackage{dsfont}
\usepackage{url}

\usepackage{color}

\newcommand{\auaff}{
Institute for physics and astronomy,
Aarhus University,
8000-DK Aarhus C,
Denmark
}
\newcommand{\nbiaff}{
Center for Quantum Devices,
Niels Bohr Institute,
University of Copenhagen,
2100-DK Copenhagen,
Denmark
}
\newcommand{\kvantifyaff}{
Kvantify Aps, 
DK-2300 
Copenhagen S, 
Denmark
}

\begin{document}
\title{Scheme for parity-controlled multi-qubit gates with superconducting qubits}
\author{Kasper Sangild Christensen}
\affiliation{\auaff}
\affiliation{\nbiaff}
\author{Nikolaj Thomas Zinner}
\affiliation{\auaff}
\affiliation{\kvantifyaff}
\author{Morten Kjaergaard}
\affiliation{\nbiaff}

\begin{abstract}
    Multi-qubit parity measurements are at the core of many quantum error correction schemes.
    Extracting multi-qubit parity information typically involves using a sequence of multiple two-qubit gates.
    In this paper, we propose a superconducting circuit device with native support for multi-qubit \emph{parity-controlled gates} (PCG). 
    These are gates that perform rotations on a parity ancilla based on the multi-qubit parity operator of adjacent qubits, and can be directly used to perform multi-qubit parity measurements.
    The circuit consists of a set of concatenated Josephson ring modulators  and effectively realizes a set of transmon-like qubits with strong longitudinal nearest-neighbor couplings.
    PCGs are implemented by applying microwave drives to the parity ancilla at specific frequencies.
    We investigate the scheme's performance with numerical simulation using realistic parameter choices and decoherence rates, and find that the device can perform four-qubit PCGs in 30\si{\nano\second} with process fidelity surpassing $99\%$.
    Furthermore, we study the effects of parameter disorder and spurious coupling between next-nearest neighboring qubits.
    Our results indicate that this approach to realizing PCGs constitute an interesting candidate for near-term quantum error correction experiments.
    
\end{abstract}

\maketitle

\section{Introduction}
The capabilities of quantum computers and simulators have been consistently improving over the last decades, and experiments controlling a vast number of quantum degrees of freedom have been performed on a variety of platforms \cite{Arute2019,Zhang2017,King2022,Zhong2020,Graham2022,Semghini2021}.
Despite this impressive progress, significant challenges remain to be addressed in demonstrating large-scale error-corrected quantum computing.
Quantum error correction (QEC) uses many physical qubits to encode a single logical qubit so that errors can be identified and corrected before they spread and corrupt the computation \cite{Terhal2015}.
A central requirement for many QEC schemes is that operations on the physical qubits that make up the logical qubit can be implemented with sufficiently low errors, the so-called \emph{threshold} for a given QEC code \cite{Aharonov2008,Knill1998}.
Depending on the specific code, the threshold value varies \cite{Terhal2015,Roffe2019,Campbell2017,Nickerson2016}, and each code comes with its own set of requirements on connectivity, gates required and strategies for ensuring fault tolerance.
Several experiments have recently demonstrated quantum error correction using physical qubits beyond (or close to) the relevant thresholds \cite{Acharya2022,Krinner2022,RyanAnderson2021,Ofek2016}.

One of the most promising protocols for quantum error correction is the \emph{surface codes}, due to their lenient demand on threshold error rate and nearest-neighbor connectivity \cite{Fowler2012,Kitaev2003,Bravyi1998,BonillaAtaides2021,Terhal2015, Dennis2002}.
Here logical qubits are encoded on a two-dimensional lattice of qubits, and errors can be detected by measuring the parity of four adjacent qubits, i.e., extracting expectation values of operators of the form $Z_1Z_2Z_3Z_4$ and $X_1X_2X_3X_4$, where $X_i$ and $Z_i$ are Pauli operators acting on qubit $i$.
These operators are the \emph{stabilizers} of the quantum error correcting code and are equivalent to finding the parity (in an appropriate basis) of a collection of qubits \cite{Gottesman1997}.
Reliable measurement of multi-qubit parity operators is thus a central feature of implementing fault-tolerant quantum computing with qubits \cite{Royer_2018,Hilder2022}.
For several quantum computing platforms, the error rates of state-of-the-art operations are (or close to) being limited by the decoherence of the qubits \cite{Sung2021,Hendrickx2021,Bruzewicz2019}.
Thus, to further reduce error rates towards implementing logical qubits, there are (at least) two issues to address: increasing the coherence of the qubits and improving the speed/efficiency of the operations required to extract the parity.


In the context of superconducting qubits, there are several avenues for improving the coherence of the physical qubits.
This includes  advancements in the materials and fabrication strategies used for superconducting qubits, as showcased by recent experiments using, e.g., tantalum-based transmons \cite{Place2021,Wang2022}, by optimizing the physical layout of the qubits on-chip, or by novel qubit circuit designs that come with error-suppressing properties out-of-the-box \cite{GyenisReview2021}, e.g. the 0-$\pi$ qubit \cite{Groszkowski2017,Gyenis2021}, the parity-protected qubit \cite{Larsen2020} 
or the bifluxon \cite{Kalashinkov2020}.
On the other hand, there have also been significant improvements in increasing the speed and reducing control errors on operations with superconducting qubits \cite{Kjaergaard2020}.
This work includes implementing and optimizing various types of two-qubit gates \cite{Foxen2020,Paik2016,Sheldon2016}, new types of coupling mechanisms including tunable couplers \cite{Yan2018,Stehlik2021,McKay2016,Ye2021}, multi-qubit gates \cite{Warren2022}, and new control schemes \cite{Sung2021,Rol2019,Negrineac2021}.
In the context of quantum error correction, two-qubit gates are concatenated to implement the desired parity operations, i.e., a four-qubit parity check (as in the surface code) requires concatenating four two-qubit gates \cite{Krinner2022,Acharya2022}.




In this paper, we propose an alternative scheme for efficiently measuring multi-qubit parity operators with superconducting circuits by a new coupling mechanism that enables multi-qubit \emph{parity-controlled gates} (PCG). 
The system we propose consists of $N$ qubits and a \emph{parity ancilla}, which is assumed to be initialized in its ground state $\ket{0_P}$. 
The key feature of our proposal is the ability to perform fast operations on the parity ancilla conditioned on the $N$-qubit parity operator $\bigotimes_i^N Z_i$, which allows for direct measurements of the multi-qubit parity.
The notion of PCGs also has applications beyond efficient stabilizer readout for QEC, and in particular, PCGs naturally arise in performing specific quantum simulation tasks. 
For instance, multi-qubit parity operators arise in the Kogut-Susskind Hamiltonian for $\mathbb{Z}_2$ lattice gauge theories \cite{Kogut1975, Munster2000}.
Another application of PCGs is for implementing the quantum approximate optimization algorithm \cite{Farhi2014}, which can be used to find approximate solutions to combinatorial optimization problems that are intractable on classical hardware. 
Specifically, using parity compilation \cite{Lechner2015,Ender2021}, multi-qubit parity plays a central role.
This technique has the advantage of requiring only nearest-neighbor connectivity with the added cost of encoding additional constraints in multi-qubit parity operators. 
Parity-controlled gates may prove useful for implementing precisely this type of dynamics. 

The paper is structured as follows:
In \cref{sec:parity_ctrld_gates}, we review how PCGs can be used to perform operations that are highly relevant in the context of quantum error correction, simulation, and optimization.
In \cref{sec:physical_implementation}, we propose a superconducting circuit device that implements PCGs efficiently. 
The circuit consists of multiple concatenated Josephson ring modulators (JRM) \cite{Abdo2012,Bergeal2010} and effectively realizes a system of qubits with strong longitudinal couplings to a parity ancilla.
Specifically, we consider the two- and four-qubit cases and use numerical simulation to gauge performance with realistic parameter choices. 
In \cref{sec:practical}, we investigate the impact of parameter disorder on device performance and the effect of unwanted next-nearest-neighbor couplings between qubits. 


\section{Parity controlled gates}
\label{sec:parity_ctrld_gates}
In this section, we introduce the notion of \emph{parity controlled gates}, that is, gates conditioned on whether a set of qubits have an even or odd number of $\ket{1}$'s.
We will denote the Hilbert space of the qubits as $\mathcal{H}_Q = \bigotimes \mathcal{H}_i$, where $\mathcal{H}_i$ is the Hilbert space of qubit $i$, and $\mathcal{H}_P$ is the Hilbert space of the parity ancilla. 
The total subspace is then the tensor product $\mathcal{H}_Q\otimes\mathcal{H}_P$. 
Our basic building block for multi-qubit control is the \emph{parity controlled flip} (PCF) gate, which is parameterized by a phase $\varphi$ and represented by the unitary
\begin{equation}
    U_{\text{PCF}}(\varphi) = P_+\otimes \mathds{1}_P + iP_-\otimes\left(\cos(\varphi)X_P+\sin(\varphi)Y_P\right).
    \label{eq:parity_flip_gate}
\end{equation}
Here $\mathds{1}_P,X_P,Y_P,Z_P$ are parity ancilla identity and Pauli operators and 
\begin{equation}
    P_\pm = \frac{1\pm\bigotimes_{i=1}^N Z_i}{2},
    \label{eq:parity_projector}
\end{equation}
are the projectors onto subspaces of different qubit parity.
For the remainder of this paper, we will omit explicit tensor products. 
As the name indicates, the purpose of the PCF-gate is to flip the parity ancilla conditioned on the parity of the adjacent qubits. 
The phase $\varphi$ dictates whether the flip occurs by rotating the parity ancilla around the $X$- or $Y$-axis on the Bloch sphere. 

A clear use case of the PCF-gate is a parity meter, which can be realized by applying the gate followed by a subsequent measurement of the parity ancilla.
The complete parity measurement process can be described using the Kraus operators
\begin{align}
    K_0 = P_+ \ket{0_P}\bra{0_P},\\
    K_1 = P_- \ket{1_P}\bra{0_P}.
    \label{eq:parity_measurement_kraus}
\end{align}
Through this measurement, we obtain information on the parity of the adjacent qubits. 
The PCF-gate can also implement a \emph{parity controlled phase} (PCP) gate. This is done using two subsequent PCF gates, which ensures that the parity ancilla ends its evolution in its ground state and, thus, that the system state remains in the computational subspace. 
Projected on the computational subspace, the total unitary reads
\begin{equation}
    \begin{split}
        U_{\text{PCP}}(\varphi) =& \bra{0_P}U_{\text{PCF}}(\pi)U_{\text{PCF}}(\varphi)\ket{0_P}\\
        =& P_+ + e^{i\varphi}P_-=e^{i\frac{\varphi}{2}}e^{-i\frac{\varphi}{2}\prod_i Z_i},
    \end{split}
\end{equation}
which simulates a Hamiltonian on form $H\propto \prod_iZ_i$. 
It also provides an alternative method for measuring the multi-qubit parity (see \cref{sec:pcp_parity_readout}).
As a demonstration of its utility, consider the case where the gate is applied to the product state $\ket{+}^{\otimes N}=2^{-1/2}\left(\ket{0}+\ket{1}\right)^{\otimes N}$ with $\varphi=\pi/2$. A simple calculation shows that the final state becomes (ignoring an unimportant overall phase factor)
\begin{equation}
    U_{\text{PCP}}\left(\frac{\pi}{2}\right)\ket{+}^{\otimes N}
    =
    \frac{1}{\sqrt{2}}\left(\ket{+}^{\otimes N}-i\ket{-}^{\otimes N}\right),
    \label{eq:ghz_state}
\end{equation}
which is equivalent to a $N$-qubit GHZ-state up to single qubit gates. 
This simple example elucidates the PCP-gates ability to generate a large amount of entanglement with only a few physical gates. 

In the following sections, we propose a superconducting circuit device capable of performing the PCF-gate using flux tuning and microwave control pulses. Furthermore, we demonstrate its performance through numerical simulations of the two- and four-qubit cases.

\section{Physical implementation}\label{sec:physical_implementation}
\begin{figure}
    \centering
    \includegraphics{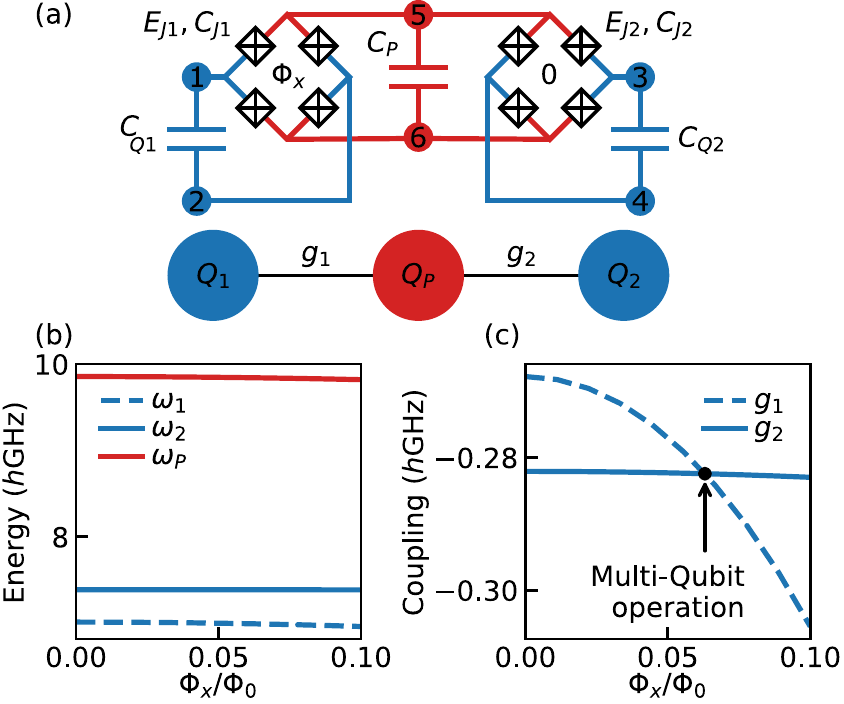}
    \caption{
    (a) 
    (Top) Diagram of the proposed circuit.
    The system consists of two capacitively shunted Josephson ring modulators that effectively realize two transmon-like qubits with a strong Cross-Kerr type coupling to a third ancillary parity qubit. 
    (b) 
    Qubit frequencies and their dependence on flux. 
    (c) 
    Flux dependence of the couplings. Multi-qubit operations are done with $g_1=g_2$.}
    \label{fig:1}
\end{figure}
We propose a realization using superconducting circuits.
The circuit consists of concatenated JRMs, a circuit element previously used in the context of quantum amplifiers \cite{Abdo2012,Ye2021}, interleaved with large shunt capacitors.
Our scheme uses the JRM modes as the physical qubits and parity ancilla. For example, the proposed circuit with two qubits can be seen in \cref{fig:1}a. Effectively, this realizes a system of qubits longitudinally connected to the central parity ancilla (see \cref{sec:circuit} for a complete circuit analysis). 
The coupling strengths and qubit frequencies depend on the flux applied through the JRM loops, and by tuning this flux, the system can be brought to a point where the coupling strengths are equal, i.e., $g_i=g$ for all $i\in \{1,...,N\}$, as demonstrated for the two-qubit case in
\cref{fig:1}b and c. By projecting the qubit and ancilla modes onto their two lowest states, we obtain the effective hamiltonian
\begin{equation}
        H_0 = -\frac{1}{2}\left(\omega_P+\Delta(Z_1,...,Z_N)\right)(Z_P-1) -\frac{1}{2}\sum_{i=1}^N\omega_iZ_i,
    \label{eq:tls_hamil}
\end{equation}
where $\omega_i$ is the frequency of the $i$'th qubit, $\omega_P$ is the frequency of the parity ancilla and 
\begin{equation}
    \Delta(Z_1,...,Z_N) = -\frac{g}{2}\sum_i^NZ_i,
    \label{eq:detuning_operator}
\end{equation}
is the detuning of the parity ancilla arising from the coupling to the adjacent qubits. 

Before we continue, it is informative to consider how this coupling scheme compares to other examples commonplace in superconducting circuits. 
The modes of the JRMs are akin to transmons \cite{Koch2007}, so it is natural to ask how our proposal differs from the highly successful transmon-based architectures commonly used in state-of-the-art quantum computing experiments. Typically, transmons couple transversely \cite{Krantz2019,Rasmussen2021}, i.e. through terms on the form $(b_1b_2^\dagger+b_1^\dagger b_2)$, where $b(b^\dagger)$ is the annihilation(creation) operator. This type of coupling can be achieved by capacitively coupling the qubits or by coupling the qubits through a coupler, e.g., a resonator \cite{Blais2004,Majer2007}. 
Furthermore, the coupling can be made flux tunable and can be entirely switched off \cite{Yan2018,Kounalakis2018}.
The main difference between traditional coupling schemes and ours is that our coupling is purely longitudinal, i.e., on the form $b_1^\dagger b_1b_2^\dagger b_2$. 
The longitudinal coupling axis allows us to have strong coupling without worrying about leakage to higher levels since $\bra{11}b_1^\dagger b_1b_2^\dagger b_2\ket{20}=0$.
The coupling arises because of the non-linearity of the Josephson inductance, the same mechanism that gives rise to the anharmonicity of transmons. 
As such, it is not surprising that the coupling strength is of the same order as the anharmonicity, typically a few hundred $h\si{\mega\hertz}$. 
This is significantly larger than the typical coupling strength between transmons, which is on the order of tens of $h\si{\mega\hertz}$.
The drawback of this large longitudinal coupling is that it cannot be switched off, even though the coupler is tunable. 
Our proposed workaround is to couple the qubits to an ancillary mode, in this case, the parity ancilla. 

In the rest of the analysis, we will work in the interaction picture given by $\ket{\psi_I}=\exp(iH_0t)\ket{\psi_S}$. The coupling strength in our scheme is large, comparable to the anharmonicity, which makes the control of individual qubits more difficult. Thus, our first order of business is to investigate how single qubit control can be implemented in the rotating frame. Consider a microwave drive applied to qubit $i$ at frequency $\omega_i$. The drive couples to the qubits charge degree of freedom $Q_i\propto Y_i$, and in the Schrödinger picture, the driving Hamiltonian reads
\begin{equation}
    H_{d,S}(t) = \Omega(t)\sin\left(\omega_i t - \varphi\right)Y_i,
\end{equation}
where $\Omega(t)$ is the pulse envelope and $\varphi$ is the phase shift of the drive.
In the interaction picture, this becomes
\begin{equation}
    H_{d,I}(t) = -\frac{\Omega(t)}{2}\left(e^{-i\left(\varphi-g(Z_P-1)\right)}\ket{0_i}\bra{1_i}+h.c.\right),
    \label{eq:h_sq}
\end{equation}
after employing rotating wave approximation (RWA).
In general, this operator depends on the state of the parity ancilla, which complicates single qubit control. 
This is the price of having strongly coupled qubits. In our scheme, the problem is circumvented by only applying single qubit gates when the parity ancilla is known to be in its ground state. 
In this case \cref{eq:h_sq} reduces to 
\begin{equation}
    \bra{0_P}H_{d,I}(t)\ket{0_P} = -\frac{\Omega(t)}{2}\left(\cos(\varphi)X_i+\sin(\varphi)Y_i\right),
    \label{eq:h_sq_0P}
\end{equation}
which can perform arbitrary single-qubit rotations. 
In summary, the system effectively decouples when the parity ancilla is in its ground state, allowing for individual qubit control and readout. 

In the rest of this section, we investigate how parity-controlled gates can be performed by applying a microwave drive to the parity ancilla.

\subsection{Two qubit case}\label{sec:two_qb}
\begin{figure}
    \centering
    \includegraphics{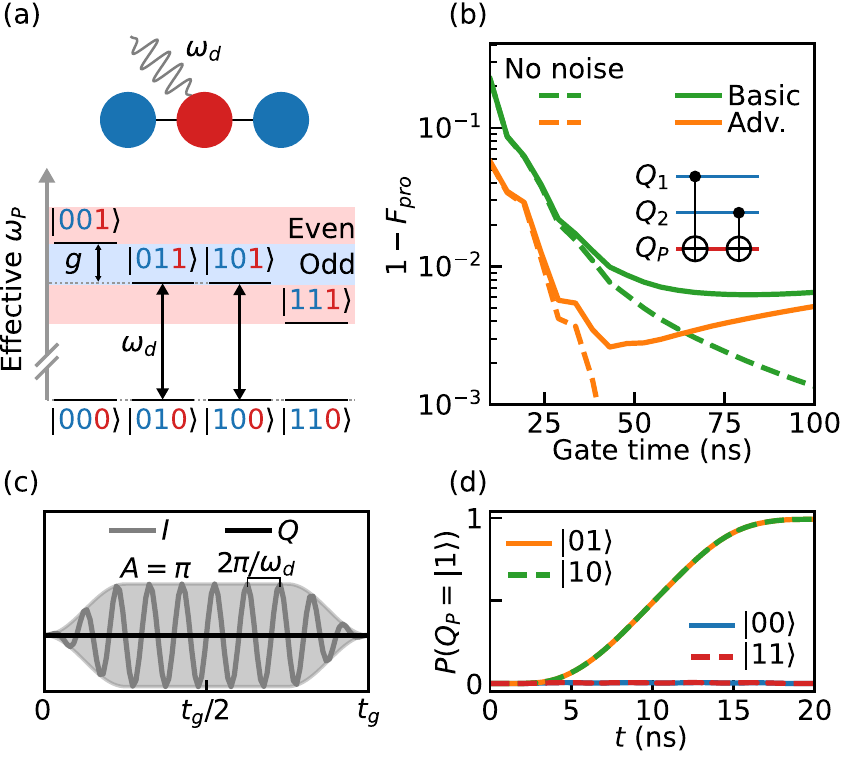}
    \caption{
    (a)
    Drive applied to the parity ancilla qubit. The drive is resonant with the $\ket{010}\leftrightarrow\ket{011}$ and $\ket{100}\leftrightarrow\ket{101}$ transitions, as indicated on the level diagram.
    (b)
    Process fidelity error as a function of gate time between the simulated process and the ideal PCF gate. 
    The solid lines are simulations including noise in the form of single qubit relaxation and dephasing with rates $T_\phi = T_1 = 40 \si{\micro\second}$.
    The advanced pulse scheme includes corrections due to AC-Stark-induced energy shifts.
    (c)
    The shape of the applied $\pi$-pulse applied to the parity ancilla qubit. 
    (d)
    Excited state probability of the parity ancilla during a single PCF gate for different qubit states. 
    }
    \label{fig:2}
\end{figure}
We now focus on the $N=2$ case shown in \cref{fig:1}a. 
The coupling strength and, to a lesser extent, the qubit frequencies depend on the externally applied flux, as shown in \cref{fig:1}b and c, and the system is operated at a bias flux so that $g_1=g_2=g$.
We apply a drive to the parity ancilla at the frequency $\omega_P$, which is resonant with the parity ancilla when $\Delta_P(Z_1, Z_2)=0$ corresponding to an odd parity configuration of the qubits as shown on \cref{fig:2}a. 
In the Schrödinger picture, the driving Hamiltonian reads
\begin{equation}
    H_{d,S}(t) = \Omega(t)\sin\left(\omega_P t + \varphi\right)Y_P.
\end{equation}
Again we will switch to the interaction frame and eliminate counter-rotating terms using RWA.
The resulting Hamiltonian is
\begin{equation}
    H_{d, I}(t) = -\frac{\Omega(t)}{2}\left(
    e^{-i\left(\varphi+g\left(Z_1+Z_2\right)t/2\right)}\ket{0_P}\bra{1_P}+h.c.
    \right).
    \label{eq:two_qubit_hamil_rotatin}
\end{equation}
Note that in the case of odd parity, we have $(Z_1+Z_2)\ket{\psi_\text{odd}} = 0$.
This time we will also assume that $|\Omega / 2|\ll |g|$ such that terms rotate with frequencies of $|g|$ or higher can be eliminated.
The driving Hamiltonian then becomes
\begin{equation}
    H_{d,I}(t) = -P_-\frac{\Omega(t)}{2}
    \left(
    \cos(\varphi)X_P+\sin(\varphi)Y_P
    \right),
    \label{eq:h_dI_after_rwa_2qb}
\end{equation}
with $P_\pm$ as defined in \cref{eq:parity_projector}.
For a fixed phase $\varphi$ the time evolution operator at time $T$ becomes 
\begin{equation}
\begin{split}
 U &= e^{-i\int_0^TH_{d,I}(t)dt} \\
 &= P_++P_-e^{i\left(\cos(\varphi)X_P+\sin(\varphi)\right)\int_0^T\frac{\Omega(t)}{2}dt}.
\end{split}
\end{equation}
After a $\pi$-pulse, where $\int_0^T\Omega(t)dt=\pi$, we recover the PCF unitary from \cref{eq:parity_flip_gate}.
Until now, we have assumed that off-resonant terms have no effect and can be eliminated using RWA. 
A more realistic model can be built by accounting for the AC-Stark shifts caused by the driving by using time-averaging techniques \cite{James2007}.
The AC-stark-induced errors can be corrected by slightly modifying the pulse (see \cref{sec:ac-stark}).
Undoubtedly, more sophisticated pulse engineering techniques can provide better results. Still, this relatively simple correction is sufficient to reach fidelities above 99\% in our case with reasonable circuit parameters and decoherence times.

To gauge the feasibility of the proposed protocol, we have carried out numerical simulations of the driven system by numerical integrating the Gorini–Kossakowski–Sudarshan–Lindblad (GKSL) equation \cite{Breuer2007}
\begin{equation}
    \dot{\rho} = -i \left[ H_{d, I}, \rho \right] +\sum_i \gamma_i\left(L_i^\dagger \rho L_i - \frac{1}{2}\{L_i^\dagger L_i, \rho\} \right),
\end{equation}
where $\rho$ is the system density matrix, and $L_i$ are the systems Lindblad operators with rates $\gamma_i$. 
We note in passing that we use SciPy \cite{Virtanen2020} and NumPy \cite{harris2020} for all numerical experiments.
Transmon-like modes such as the ones used in this scheme are weakly anharmonic. 
As such, we model the parity ancilla as a three-level system with anharmonicity $\alpha_P$ for simulation. 
We have used parameters $g=-250h\si{\mega\hertz}$, $\alpha_P=-100h\si{\mega\hertz}$ consistent with realistic hardware parameters (see \cref{sec:circuit}). 
The system relaxes to the ground state for low temperatures, so we model relaxation through the annihilation operator $L_{\text{rel},i} = b_i$, and dephasing using the number operators of each mode $L_{\phi,i} = b_i^\dagger b_i$.
We use relaxation time $T_1=40\si{\micro\second}$ and pure dephasing time $T_\phi=40\si{\micro\second}$.
We use a Tukey window function for the pulse shape and set $\varphi=0$ as shown in \cref{fig:2}b.
As expected, the parity ancilla is only excited when the qubits are in state $\ket{01}$ or $\ket{10}$ as shown in \cref{fig:2}d.
By subsequently measuring the parity ancilla, we realize the parity measurement operation given in \cref{eq:parity_measurement_kraus}.
As a measure of process quality, we use the process fidelity \cite{Gilchrist2005} defined as
\begin{equation}
    F_{\text{pro}}\left(\mathcal{E}_{\text{sim}}, \mathcal{E}_{\text{ideal}}\right) = F\left(\rho_{\mathcal{E},\text{sim}},\rho_{\mathcal{E},\text{ideal}}\right),
\end{equation}
here $\mathcal{E}_{\text{sim}}$($\mathcal{E}_{\text{ideal}}$) is the simulated (ideal) process, $\rho_{\mathcal{E}}$ is the Choi density matrix representation \cite{Choi1975,Jiang2013} of the process $\mathcal{E}$ and $F(\rho, \sigma)=\left(\text{tr}\sqrt{\sqrt{\rho}\sigma\sqrt{\rho}}\right)^2$ is the fidelity \cite{Nielsen2011}. The simulations show process fidelities above 99\% can be achieved in roughly 25$\si{\nano\second}$ (see \cref{fig:2}c). For low gate times, the error is dominated by coherent errors, such as leakage and dynamical phase error, while relaxation and incoherent dephasing limit fidelity for longer gate times. 

\subsection{Four qubit case}\label{sec:four_qb}
\begin{figure}
    \centering
    \includegraphics{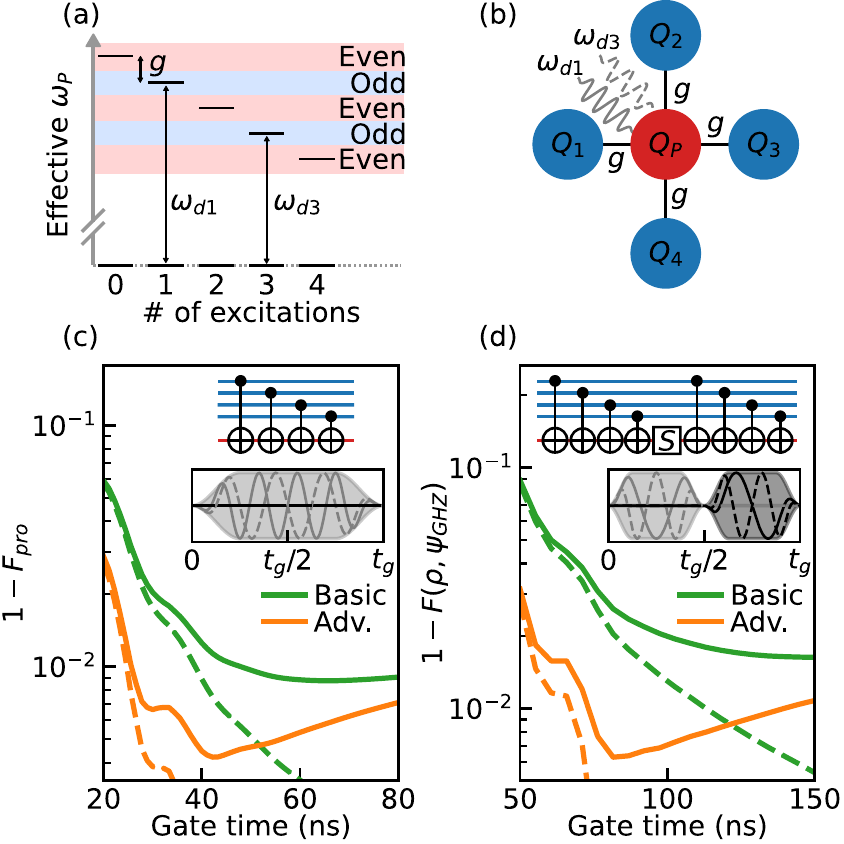}
    \caption{
    (a)
    Level diagram of a four-qubit plaquette. 
    A two-tone drive is applied to the parity ancilla, as shown. The drive is resonant if the qubits have an odd number of excitations.
    (b)
    Plaquette configuration with four qubits and a parity ancilla. The four qubits couple longitudinally to the central parity ancilla with equal strength.
    (c)
    Process fidelity as a function of gate time for the four qubit PCF gate.
    The inset shows the pulse shape used for the (basic) simulation. 
    The advanced simulation includes corrections to AC-Stark-induced errors.
    The solid lines are simulated with decoherence in the form of single qubit dephasing and relaxation. The rates used are $T_1=T_\phi=40\si{\micro\second}$.
    (d)
    Simulation of the GHZ-factory circuit. The inset shows the ideal circuit and the pulse sequence used to realize it. The plot shows the state fidelity of the output state with the target GHZ state.
    }
    \label{fig:3}
\end{figure}
An important feature of our scheme is its extension to two dimensions, which we do by connecting two additional qubits to the parity ancilla, as shown in \cref{fig:3}b.
As in the two-qubit case, multi-qubit operations are performed by applying a drive to the parity ancilla. In this case, we use a two-tone drive at frequencies $\omega_\pm=\omega_P\pm g$, as shown on \cref{fig:3}a. 
In the Schrödinger picture, the driving Hamiltonian becomes
\begin{equation}
    H_{d,S}(t) = \Omega(t)\left(\sin\left(\omega_+ t - \varphi\right)+\sin\left(\omega_- t - \varphi\right)\right)Y_P.
\end{equation}
Note that one of the tones is resonant in the case where $\Delta_P(Z_1,...,Z_4) = -g$ and the other is when $\Delta_P(Z_1,...,Z_4) = g$. 
\cref{eq:detuning_operator} and \cref{fig:2}a shows that these two cases correspond to cases where there are either 1 or 3 excitations amongst the qubits, i.e., the states with odd parity. 
Switching to the interaction picture and employing RWA, we obtain 
\begin{equation}
\begin{split}
    H_{d, I}(t) =& -\frac{\Omega(t)}{2}
    e^{-i\left(\varphi+gt\left(1+\sum_{i=1}^4 Z_i/2\right)\right)}\ket{0_P}\bra{1_P}\\
    &-\frac{\Omega(t)}{2}
    e^{-i\left(\varphi-gt\left(1-\sum_{i=1}^4 Z_i/2\right)\right)}\ket{0_P}\bra{1_P}\\
    &+h.c.\\
    \approx & -P_-\frac{\Omega(t)}{2}
    \left(
    \cos(\varphi)X_P+\sin(\varphi)Y_P,
    \right)
\end{split}
\label{eq:h_dI_after_rwa_4qb}
\end{equation}
where we have employed RWA a second time in the last line to eliminate terms rotating with frequencies $\sim g$.
\cref{eq:h_dI_after_rwa_4qb} is the four-qubit version of \cref{eq:h_dI_after_rwa_2qb}, and the PCF unitary can be realized using a single $\pi$-pulse following the same arguments as before.

As with the two-qubit case, we have conducted numerical simulations of the PCF-gate.
We use a coupling strength of $g=-200h\si{\giga\hertz}$ consistent with realistic hardware parameters (see \cref{sec:circuit}), and an anharmonicity of $-100h\si{\giga\hertz}$.
\cref{fig:3}c shows the process fidelity of the simulated 
PCF-gate with and without AC-stark compensation. 
We achieve process fidelities above 99\% in roughly 30$\si{\nano\second}$.
Furthermore, we have carried out simulations of the PCP-gate.
Specifically, we simulate its performance as a GHZ-state factory, as shown on \cref{fig:3}d, where we apply the simulated PCP gate to create the GHZ-like state described in \cref{eq:ghz_state}.
Here we can initialize the GHZ state with 99\% fidelity in roughly $75\si{\nano\second}$. 

\section{Practical considerations}
\label{sec:practical}
\begin{figure}
    \centering
    \includegraphics{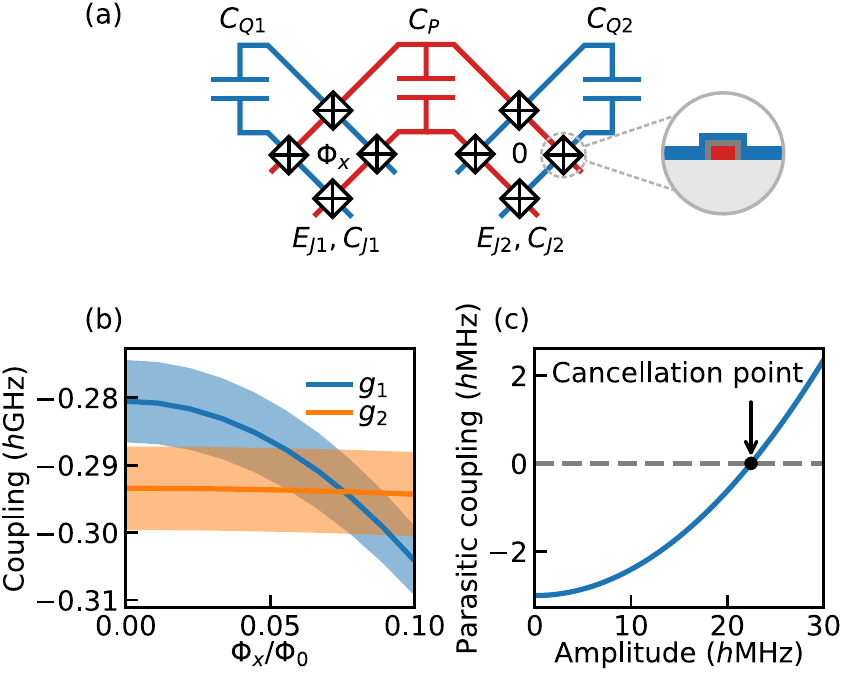}
    \caption{
    (a) 
    Sketch illustrating the proposed physical layout of the circuit.
    Using cross-over junctions, as shown on the inset, we eliminate the need for air bridges. 
    (b)
    A plot of the coupling strengths illustrating the effects of parameter disorder. The mean and standard deviation of 200 realizations of the circuit with randomized Josephson energies have been calculated. Each Josephson energy is picked from a normal distribution with mean at the ideal energy and a standard deviation of 10\%.
    (c)
    Cancellation of the parasitic coupling between qubits through off-resonant driving of the parity ancilla. 
    }
    \label{fig:4}
\end{figure}
We propose a physical layout using crossover junctions as indicated on \cref{fig:4}a. This compact design eliminates the need for air bridges and may prove advantageous for scalability. 
Furthermore, the design is easily adapted to two dimensions.

So far, our analysis has assumed that junctions belonging to the same JRM have identical $E_J$'s. 
In practice, fabrication imperfections will almost always allow for some degree of parameter disorder among the junctions. 
The disorder results in transverse couplings between qubits and parity ancilla. 
The coupling is relatively small (on average $\sim 50h\si{\mega\hertz}$) compared to the qubit-parity ancilla detuning and can therefore be treated as a perturbation.
Ultimately, the parameter disorder amounts to a shift in the coupling strength between qubits and parity ancilla as shown on \cref{fig:4}b. 
Importantly, the shift is small enough that the system can be tuned such that $g_1=g_2$, even in the presence of disorder.

Another issue common to qubit couplers in superconducting circuits is the unwanted coupling between non-neighboring qubits.
This issue is also present in our scheme, although its origin differs from typical superconducting couplers.
The coupling arises from higher order correction to the Hamiltonian and is on the form $g_{12}Z_1Z_2$ where $g_{12} \propto g_1g_2/ (2\omega_P)$.
Although the parasitic coupling is suppressed by $|g/\omega_P|\ll 1$, it is still significant due to the large value of $g$.
This coupling can be compensated for by applying an off-resonant drive to the parity ancilla, as demonstrated on \cref{fig:4}c where a drive at frequency $\omega_P-g/2$ has been applied. 
By tuning the amplitude, the parasitic coupling is canceled by the driving-induced AC-Stark shift. 
Similar schemes have been proposed to compensate for unwanted couplings in other coupler designs \cite{Wei2021}. 
Additionally, we note that the parasitic coupling scales as $g^2$, whereas the gate time scale as $g^{-1}$. 
Thus, by reducing the coupling strength between qubits and parity ancilla, the parasitic coupling is suppressed while only increasing gate times moderately.

\section{Conclusion}
In this paper, we have presented a novel scheme for implementing parity-controlled gates in superconducting circuits. 
The scheme relies on a strong longitudinal coupling between qubits and an ancillary parity qubit. 
We have performed numerical simulations of the scheme with realistic circuit and decoherence parameters for the two- and four-qubit cases.
In both cases, we achieve process fidelities above 99\% in the order of tens of nanoseconds. 
Furthermore, we examine the effects of parameter disorder in the junction energies and find that this only slightly modifies the qubit-ancilla coupling.
Our simulations indicate that the tunability of the proposed device can compensate for these effects. 
We have also demonstrated how unwanted parasitic coupling between qubits may be canceled by applying an off-resonant drive to the parity ancilla.

All in all, the proposed scheme natively implements features that are highly desirable in many areas of quantum computing and presents an exciting test-bed for surface codes and specific quantum simulation and optimization schemes. 

\section{Acknowledgements}
The authors thank András Gyenis for enlightening discussions on the proposed device.
KSC and NTZ acknowledge support from the Independent Research Fund Denmark.
MK gratefully acknowledges support from the Danish National Research Foundation, the Danish Council for Independent Research | Natural Sciences, Villum Foundation (grant 37467) through a Villum Young Investigator grant.

\appendix

\section{Parity readout using the PCP-gate}
\label{sec:pcp_parity_readout}
In the main text, we proposed performing a PCF gate followed by measurement of the parity ancilla to measure the multi-qubit parity operator $\prod_iZ_i$.
In this section, we propose an alternate scheme that instead involves measuring one of the qubits, say qubit 1, for concreteness. 
This may prove advantageous since the system effectively decouples while the parity ancilla is in its ground state, and it is straightforward to perform qubit measurements in this case. 
This alternate readout protocol involves a single PCP gate sandwiched between single-qubit gates on qubit 1. 
Specifically:
\begin{equation}
    U_{1} = 
    H_1S_1^\dagger U_{PCP}\left(\frac{\pi}{4}\right)H_1,
\end{equation}
where $H_1=(X_1+Z_1)/\sqrt{2}$ is the Hadamard gate and $S_1 = \text{diag}(1, i)$ the phase gate. 
It can be seen that this unitary allows for parity measurements by considering 
$U_1^\dagger Z_1U_1 = \prod_iZ_i$.
Thus, after a single $U_1$ gate, the parity information is encoded in qubit one and can be measured.
The measurement may change the state of qubit 1; however, this can be undone by subsequently applying $U_1^\dagger$.

\section{Circuit}\label{sec:circuit}
\begin{figure}
    \centering
    \includegraphics{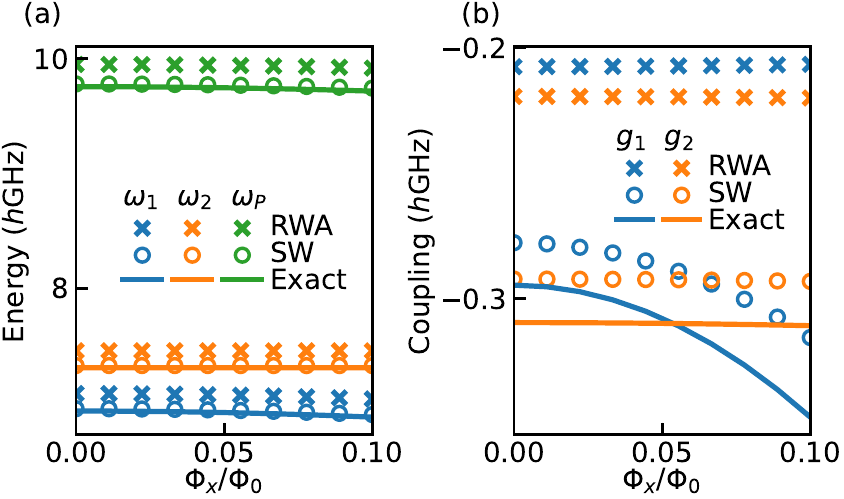}
    \caption{(a) Frequencies of qubits and parity ancilla computed using naive RWA, second order perturbation theory using Schrieffer wolf transformation (SW) and exact diagonalization. (b) Comparison of coupling strengths between RWA, perturbative and exact diagonalization approaches.}
    \label{fig:exact_diagonalization}
\end{figure}
The circuit we propose is shown in \cref{fig:1}a.
The circuit consists of three large capacitors coupled through four Josephson junctions in a ring configuration.
This section shows how our system reduces to a chain of three modes with strong, independently tunable longitudinal couplings between neighboring qubits.
For convenience, we will work in units where $2e = \hbar = 1$.

The first step is to define the coordinates in terms of the node fluxes of the circuit.
We write this transformation as
\begin{equation}
    \begin{pmatrix}
        \phi_1 \\ \phi_2 \\ \phi_P \\ \theta_1\\ \theta_2
    \end{pmatrix}
    =
    \frac{1}{2}
    \begin{pmatrix}
        1 & -1 & 0  & 0  & 0  & 0  \\
        0 & 0  & 1  & -1 & 0  & 0  \\
        0 & 0  & 0  & 0  & 1  & -1 \\
        1 & 1  & 0 & 0 & -1  & -1  \\
        0 & 0  & 1  & 1  & -1 & -1
    \end{pmatrix}
    \begin{pmatrix}
        \varphi_1 \\
        \varphi_2 \\
        \varphi_3 \\
        \varphi_4 \\
        \varphi_5 \\
        \varphi_6
    \end{pmatrix},
\end{equation}
where the $\varphi_i$'s are the node fluxes, and $\phi_i$ and $\theta_i$ constitute the chosen coordinates.
In these coordinates, the potential energy takes the form
\begin{equation}
    \begin{split}
        V = \sum_{i=1}^2-4E_{Ji}&\left(\cos{\phi_i}\cos{\phi_P}\cos{\theta_i}\cos{\frac{\phi_{xi}}{4}} \right.\\
        &-\left.\sin{\phi_i}\sin{\phi_P}\sin{\theta_i}\sin{\frac{\phi_{xi}}{4}}    \right),
    \end{split}
\end{equation}
where we have defined the normalized external flux $\phi_{x1} = 2\pi\Phi_x/\Phi_0$.
The kinetic energy is
\begin{equation}
    K = E_{CP}q_P^2+\sum_{i=1}^2\left(E_{Ci}q_i^2+\tilde{E}_{Ci}Q_i^2\right),
\end{equation}
where we have defined the canonical conjugate momenta $\{\phi_i, q_i\}=\{\theta_i, Q_i\}=\{\phi_P, q_P\}=1$, where $\{\cdot, \cdot\}$ denotes the Poisson bracket. 
The charging energies are given by
\begin{align}
        E_{Ci} =& \frac{1}{8(C_{Qi}+C_{Ji})},\\
        E_{CP} =& \frac{1}{8(C_P+C_{J1}+C_{J2})},\\
        \tilde{E}_{Ci} =& \frac{1}{8C_{Ji}},\\
\end{align}
where $i \in \{1, 2\}$.
The device is operated near $\phi_{xi}=0$, meaning that the potential has a minimum at $\phi_i=\theta_i=\phi_P=0$. 
Therefore we assume that the wavefunction is reasonably localized around this point, and we can approximate the potential by its Taylor expansion. 
The system then reduces to a system of coupled anharmonic oscillators and is quantized by introducing the ladder operators $b_\pm = \left(\zeta^{-1/2}\phi\mp i\zeta^{1/2}q\right)/\sqrt{2}$, where $\zeta = \sqrt{E_C/(2E_J)}$ is the modes characteristic impedance.
After expanding, the Hamiltonian becomes
\begin{equation}
    \begin{split}
        H =& 
        E_{CP}q_P^2+4\sum_{i=1}^2c_iE_{Ji}\left(\frac{\phi_P^2}{2}-\frac{\phi_P^4}{24}\right)\\
        & 
        +\sum_{i=1}^2\left(E_{Ci}q_i^2+4c_iE_{Ji}\left(\frac{\phi_i^2}{2}-\frac{\phi_i^4}{24}\right)\right)\\
        & 
        +\sum_{i=1}^2\left(\tilde{E}_{Ci}Q_i^2+4c_iE_{Ji}\left(\frac{\theta_i^2}{2}-\frac{\theta_i^4}{24}\right)\right)\\
        &-\sum_{i=1}^2c_iE_{Ji}\left(\phi_i^2\phi_P^2+\phi_{i}^2\theta_i^2+\phi_{P}^2\theta_i^2\right)\\
        &+4\sum_{i=1}^2s_iE_{Ji}\phi_i\theta_i\phi_P,
    \end{split}
    \label{eq:hamil_taylor}
\end{equation}
where we have introduced $c_i = \cos\left(\phi_{xi} / 4\right)$ and $s_i = \sin\left(\phi_{xi} / 4\right)$ to ease notation. 
In the following, we will work with a basis of eigenstates for the quadratic Hamiltonian and treat off-diagonal, which consists only of cubic and quartic terms, as a perturbation. 
We denote the diagonal part of the Hamiltonian as $H_D=\sum_\alpha E_\alpha \ket{E_\alpha}\bra{E_\alpha}$ and the off-diagonal as $H_X$. 
Projecting onto the relevant subspace, where the parity ancilla and qubits are restricted to their two lowest eigenstates and the $\theta$-modes are in their ground states, yields a Hamiltonian on the form
\begin{equation}
    \begin{split}
        H_D = -\frac{1}{2}\left(\omega_P-\frac{g_1}{2}Z_1-\frac{g_2}{2}Z_2\right)(Z_P-1) -\frac{1}{2}\sum_{i=1}^2\omega_iZ_i,        
    \end{split}
\end{equation}
with the unperturbed frequencies $\omega_P$ and $\omega_i$, and unperturbed coupling strengths $g_i$.
The primary mechanism behind the strong longitudinal coupling is the quartic interaction since the diagonal part of $\phi_P^2\phi_i^2\propto (b_P^\dagger b_P+1)(b_{\phi i}^\dagger b_{\phi i}+1)$.
However, the off-resonant interactions of $H_X$ introduce energy shifts which can be approximated using standard perturbation techniques. 
To second order in $H_X$, the corrections are
\begin{equation}
    E^{(2)}_\alpha = \sum_{\beta\neq\alpha}\frac{\left|\bra{E_\alpha}H_X\ket{E_\beta}\right|^2}{E_\alpha-E_\beta},
    \label{eq:perturbation}
\end{equation}
where the $\beta$ runs over all states, not only those in the relevant subspace.
These energy shifts modify the coupling strengths and mode frequencies. 
This is shown on \cref{fig:exact_diagonalization}, which demonstrates a significant discrepancy between the naive RWA approach, where the off-diagonal terms are eliminated, and the exact diagonalization of the system. 
The second-order perturbative approach is much closer to the exact diagonalization and captures the coupling strength's flux dependence.
Furthermore, the quartic interaction terms introduce a small parasitic coupling between the qubits, which can be canceled by applying an off-resonant driving, as described in the main text.

In our numerical simulations, we are only interested in the lowest $N$ levels of each mode, where $N=3$ for the parity ancilla, $N=2$ for the qubits, and $N=1$ for the $\theta$ modes. 
We compute the energy shifts of each level according to \cref{eq:perturbation} using the lowest $N+2$ levels of each mode, thus including all shifts due to interactions with higher levels. 
We wish to choose the external flux biases so that $g_i = g$ for all $i$ by numerically minimizing $\text{Var}(g_i)$.
Although it is possible to diagonalize the system for two qubits numerically, this becomes unfeasible in the four-qubit case for the hardware used for this work since the total dimension of the Hilbert space becomes $5\cdot 4^4 \cdot 3^4 \sim 10^5$ dimensional.
As such, we rely on the perturbative approach for diagonalization. 

For the simulations of the two-qubit system, we have used the capacitances
$C_1 = C_2 = 22\si{\femto\farad}$, 
$C_P = 19 \si{\femto\farad}$ and 
$C_{J1} = C_{J2} = 4\si{\femto\farad}$, 
together with junction energies
$E_{J1}=10h\si{\giga\hertz}$ and 
$E_{J2}=11h\si{\giga\hertz}$. 
This give an optimal flux bias of $\Phi_{x1}/\Phi_0=\phi_x/(2\pi)\approx 0.06$ and $\Phi_{x2}=0$ with a coupling strength of roughly $280h\si{\mega\hertz}$, and frequencies ranging from $7.0-7.4h\si{\giga\hertz}$ for the qubit and $9.8h\si{\giga\hertz}$ for the parity ancilla, as seen on \cref{fig:1}c.
For the four-qubit case, we use Josephson energies 
$E_{J1}=E_{J2}=13h\si{\giga\hertz}$ and
$E_{J3}=E_{J4}=12h\si{\giga\hertz}$ 
with capacitances
$C_1 = C_3 = C_P = 13\si{\femto\farad}$, 
$C_2 = C_4 = 15\si{\femto\farad}$ and 
$C_{Ji} = 4\si{\femto\farad}$ for $i=1,...,4$.
In this case, the qubit frequencies are all between $8.8h\si{\giga\hertz}$ and $9.6h\si{\giga\hertz}$, while the parity ancillas frequency is $15h\si{\giga\hertz}$. 
The coupling strength between qubits and parity ancilla is $214h\si{\mega\hertz}$.
In both the two- and four-qubit cases, the theta modes have frequencies between $19.4h\si{\giga\hertz}$ and $20.6h\si{\giga\hertz}$.

\section{Accounting for AC-stark shifts}\label{sec:ac-stark}
The rotating frame Hamiltonian given by \cref{eq:h_dI_after_rwa_2qb} only holds in the limit where $|\Omega / g|\rightarrow 0$. 
Furthermore, when driving the $\ket{0}\xrightarrow{}\ket{1}$ transition of a weakly anharmonic qubit, such as the transmon-like qubits of our system, one also drives the $\ket{1}\xrightarrow{}\ket{2}$ transition off-resonantly.
The off-resonant driving of specific transitions leads to AC-stark shifts of the levels, ultimately leading to phase errors if not accounted for.

In this section, we demonstrate a simple technique for suppressing this type of error using time-averaging techniques. 
Intuitively, this can be considered a more sophisticated version of RWA. 
To gauge the effect of leakage, we model the parity ancilla as an anharmonic oscillator with anharmonicity $\alpha$. 
This means modifying \cref{eq:tls_hamil}:
\begin{equation}
    \begin{split}
    H_0 = &
    -\sum_i\omega_iZ_i/2\\
    &+
    \left(\omega_P+\Delta_P\right)b^\dagger_P b_P+\frac{\alpha}{2}b^\dagger_P b_P(b^\dagger_P b_P-1),
\end{split}
\end{equation}
where $b^\dagger_P$ and $b_P$ are the parity ancilla creation and annihilation operators.
Since there are only longitudinal couplings and no driving on the qubits, it is not necessary to include the higher energy levels of the qubit degrees of freedom.
We show that the AC-stark-induced errors can be accounted for by changing the phase of the driving pulse. 
It is convenient to define the projection onto the relevant subspace
\begin{equation}
    P = P_+\ket{0_P}\bra{0_P}+P_-\left(\ket{0_P}\bra{0_P}_P+\ket{1_P}\bra{1_P}\right).
    \label{eq:relevant_projector}
\end{equation}
If the qubits are in an even parity state, this projects onto the ground state of the parity ancilla, and if the qubits are in an odd parity state, it projects onto the two lowest states instead.

\subsection{Two qubit case}\label{subsec:ac-stark_2qb}
We apply a drive to the parity ancilla coupled to two qubits at frequency $\omega_d$. In the Schrödinger picture, the Hamiltonian reads
\begin{equation}
    H_{S} = H_0+H_{d},
\end{equation}
is the free Hamiltonian of the system and 
\begin{equation}
    H_{d} = i\Omega(t) \sin(\omega_dt - \phi(t))(b^\dagger - b).
\end{equation}
We now switch to a frame defined by
\begin{equation}
    \ket{\psi(t)}_R = V(t)\ket{\psi(t)}_S, 
\end{equation}
with $V(t)=\exp\left(i\int_0^tH_R(t)dt\right)$ and 
\begin{equation}
    H_R(t) = H_0+H_{ac}(t),
\end{equation}
where
\begin{equation}
    H_{ac} =\delta_P(t) b^\dagger_P b_P -\frac{\delta_Q(t)}{2}\sum_iZ_i,
    \label{eq:H_ac_two_qubit}
\end{equation}
is introduced to cancel AC-stark shifts induced by the driving term.
Crucially, the rotating frame Hamiltonian depends only on diagonal single qubit operators, and we can therefore switch between this frame and the lab frame using only virtual rotation. It is also worth noting that in the relevant subspace, we have 
\begin{equation}
    PH_{ac}P 
    = \delta_Q\left(P_2-P_0\right)\ket{0_P}\bra{0_P}+\delta_PP_-\ket{1_P}\bra{1_P},
\end{equation}
where $P_n$ is the projector onto the qubit subspace of Hamming weight $n$.
We now choose the driving frequency as $\omega_d = \omega_P$ and the phase as $\phi(t)=\varphi-\int_0^t\delta_{P}(t')dt'$. 
Switching to the rotating frame and eliminating terms rotating with frequencies of $2|\omega_d|$, we find
\begin{equation}
\begin{split}
    H_{d, I} =& VH_dV^\dagger -H_{ac}\\
    =& \frac{\Omega (t)}{2}\left(b^\dagger (t)+b (t)\right)-H_{ac},
    \label{eq:ip_drive_hamil_ac_stark}
\end{split}
\end{equation}
where for compactness, we have defined
\begin{align}
    b(t) =e^{-i\left(\Delta_Pt+\varphi\right)}\left(\ket{0_P}\bra{1_P}+\sqrt{2}e^{-i\alpha t}\ket{1_P}\bra{2_P}\right).
\end{align}
Computing the effective time-averaged Hamiltonian \cite{James2007} and projecting onto the relevant subspace yields
\begin{equation}
\begin{split}
    PH_{eff}P =& H_{XY}(t, \varphi)\\ 
    &+\left(\delta_Q(t)+\frac{\Omega(t)^2}{4g}\right)\left(P_0-P_2\right)\ket{0_P}\bra{0_P}\\
    &-\left(\frac{\Omega(t)^2}{2\alpha}+\delta_P(t)\right)P_-\ket{1_P}\bra{1_P}.
\end{split}
\end{equation}
By setting
\begin{align}
        \delta_Q(t) =& -\frac{\Omega(t)^2}{4g},\\
        \delta_P =& -\frac{\Omega(t)^2}{2\alpha}.
\end{align}
we thus recover \cref{eq:h_dI_after_rwa_2qb}.

\subsection{Four qubit case}
In the four-qubit case, we apply a two-tone drive to the parity ancilla on the form
\begin{equation}
    H_d = i\Omega(t) \left(\sin(\omega_3t - \phi_3(t))+\sin(\omega_1t - \phi_1(t))\right)(b^\dagger - b),
\end{equation}
where we choose $\omega_d = \omega_m+\delta_m$ and $\omega_s = g+g_{ac}$, where $|g_{ac}|\ll |g|$ is a small correction to the effective coupling strengths that arise due to AC-stark shifts of off-resonantly driven transitions. 
Switching to the rotating frame defined as before but with
\begin{equation}
    H_{ac}(t)=\left(\delta_{P,1}(t)P_1+\delta_{P,3}(t)P_3\right)b^\dagger_P b_P -\frac{\delta_Q(t)}{2}\sum_iZ_i.
\end{equation}
Again we switch to a frame rotating with $H_R(t)=H_0+H_{ac}(t)$, this time choosing $\phi_i(t) = \varphi_i - \int_0^t\delta_{P,i}(t')dt'$. For the moment, $\varphi_i$ is an undetermined constant phase. In the interaction picture, the Hamiltonian becomes identical to \cref{eq:ip_drive_hamil_ac_stark} with
\begin{align}
    b(t) =2e^{-i\left(\Delta t+\varphi\right)}\cos(gt)\left(\ket{0}\bra{1}_P+\sqrt{2}e^{-i\alpha t}\ket{1}\bra{2}_P\right).
\end{align}
Here we have implicitly used that $\exp\left(i(\Delta\pm g)t+\int_0^tH_{ac}\right)\sim \exp\left(i(\Delta\pm g)t\right)$ for the purposes of time-averaging, when acting on the subspace where $\Delta\pm g\neq 0$.
In the following, we will drop the explicit time dependence for compactness. 
The effective Hamiltonian projected onto the relevant subspace becomes
\begin{equation}
\begin{split}
    PH_{eff}P =& -P_1\frac{\Omega}{2}
    \left(
    \cos(\varphi_-)X_P+\sin(\varphi_-)Y_P
    \right)\\
    & -P_3\frac{\Omega}{2}
    \left(
    \cos(\varphi_+)X_P+\sin(\varphi_+)Y_P
    \right)\\
    & +\sum_{n=0}^4\delta_{n,0}P_n\ket{0_P}\bra{0_P}
    \\
    & +\sum_{m\in \{1, 3\}} \delta_{m,1}P_m\ket{1_P}\bra{1_P}
\end{split}
\end{equation}
with
\begin{align}
    \delta_{0,0}=& -\delta_{4, 0} =2\delta_Q+\frac{\Omega^2}{3g}\\
    \delta_{1,0}=& -\delta_{3, 0} = \delta_Q+\frac{\Omega^2}{8g}\\
    \delta_{1,1}=& \delta_Q-\delta_{P,1}-\frac{\Omega^2}{8g}-\frac{\Omega^2}{2\alpha}-\frac{\Omega^2}{2\alpha-4g}\\
    \delta_{3,1}=& -\delta_Q-\delta_{P,3}+\frac{\Omega^2}{8g}-\frac{\Omega^2}{2\alpha}-\frac{\Omega^2}{2\alpha+4g}.
\end{align}
By setting 
\begin{align}
    \delta_Q=&-\frac{\Omega^2}{6g}\\
    \delta_{P,1} =& -\frac{\Omega^2}{4g}-\frac{\Omega^2}{2\alpha}-\frac{\Omega^2}{2\alpha-4g}\\
    \delta_{P,3} =& \delta_Q+\frac{\Omega^2}{8g}-\frac{\Omega^2}{2\alpha}
\end{align}
the effective Hamiltonian reads
\begin{equation}
    \begin{split}
        PH_{eff}P =& -P_1\frac{\Omega}{2}
            \left(
            \cos(\varphi_-)X_P+\sin(\varphi_-)Y_P+\frac{\Omega}{12g}
            \right)\\
            & -P_3\frac{\Omega}{2}
            \left(
            \cos(\varphi_+)X_P+\sin(\varphi_+)Y_P-\frac{\Omega}{12g}
            \right)
    \end{split}
\end{equation}
After a $\pi$-pulse, we achieve a slight variation on the PCF gate:
\begin{equation}
\begin{split}
    U_{PCF}(\varphi) =& P_++ ie^{i\chi}P_1\left(\cos(\varphi_1)X_P+\sin(\varphi_1)Y_P\right)\\
    &+ ie^{-i\chi}P_3\left(\cos(\varphi_3)X_P+\sin(\varphi_3)Y_P\right),
    \label{eq:parity_flip_gate_app}
\end{split}
\end{equation}
where $\chi = \int_0^T\Omega(\tau)^2/(24g)d\tau$. 
For parity measurements, this unitary serves precisely the same function as that of \cref{eq:parity_flip_gate} since the phases become irrelevant once the parity ancilla is measured.
The PCP may be realized using a two-pulse sequence, the first with $\varphi_1=\varphi_3=\pi$, and the second with $\varphi_1=\varphi-2\chi$ and $\varphi_3=\varphi+2\chi$.

\bibliography{references}
\bibliographystyle{unsrt}
\end{document}